\documentclass[a4paper, 10pt, conference]{ieeeconf}      
\IEEEoverridecommandlockouts                              
\usepackage{geometry}                
\usepackage{graphicx}
\usepackage{amssymb}
\usepackage{amsfonts}
\usepackage{amsmath}
\usepackage{epstopdf}
\usepackage{dsfont}

\newcommand{\mR}{\mathbb{R}}

\newcommand{\spa}{\mathrm{span \,}}

\usepackage{theorem}
{\theorembodyfont{\itshape}\newtheorem{theorem}{Theorem}}
{\theorembodyfont{\itshape}\newtheorem{proposition}[theorem]{Proposition}}
{\theorembodyfont{\itshape}\newtheorem{lemma}[theorem]{Lemma}}
{\theorembodyfont{\itshape}\newtheorem{corollary}[theorem]{Corollary}}
{\theorembodyfont{\upshape}\newtheorem{definition}[theorem]{Definition}}
{\theorembodyfont{\itshape}}
{\theorembodyfont{\upshape}\newtheorem{remark}[theorem]{Remark}}
{\theorembodyfont{\upshape}}
{\theorembodyfont{\upshape}\newtheorem{example}[theorem]{Example}}

\DeclareMathOperator{\im}{im}

\DeclareMathOperator{\diag}{diag}

\DeclareMathOperator{\trace}{trace}

\newcommand{\calE}{\ensuremath{\mathcal{E}}}

\newcommand{\calG}{\ensuremath{\mathcal{G}}}
\newcommand{\calH}{\ensuremath{\mathcal{H}}}

\newcommand{\calN}{\ensuremath{\mathcal{N}}}

\newcommand{\calV}{\ensuremath{\mathcal{V}}}

\newcommand{\hatD}{\ensuremath{\hat{D}}}
\newcommand{\hatE}{\ensuremath{\hat{E}}}

\newcommand{\hatG}{\ensuremath{\hat{G}}}

\newcommand{\hatM}{\ensuremath{\hat{M}}}

\newcommand{\hatR}{\ensuremath{\hat{R}}}

\newcommand{\bmat}{\begin{matrix}}
\newcommand{\emat}{\end{matrix}}
\newcommand{\bbm}{\begin{bmatrix}}
\newcommand{\ebm}{\end{bmatrix}}
\newcommand{\bpm}{\begin{pmatrix}}
\newcommand{\epm}{\end{pmatrix}}
\newcommand{\bse}{\begin{subequations}}
\newcommand{\ese}{\end{subequations}}
\newcommand{\beq}{\begin{equation}}
\newcommand{\eeq}{\end{equation}}
\newcommand{\ben}{\begin{enumerate}}
\newcommand{\een}{\end{enumerate}}
\newcommand{\beni}{\renewcommand{\labelenumi}{\roman{enumi}.}
\renewcommand{\theenumi}{\roman{enumi}}\begin{enumerate}}
\newcommand{\eeni}{\end{enumerate}\renewcommand{\labelenumi}{\arabic{enumi}.}
\renewcommand{\theenumi}{\arabic{enumi}}}
\newcommand{\bena}{\renewcommand{\labelenumi}{\alpha{enumi}.}
\renewcommand{\theenumi}{\alpha{enumi}}\begin{enumerate}}
\newcommand{\eena}{\end{enumerate}\renewcommand{\labelenumi}{\arabic{enumi}.}
\renewcommand{\theenumi}{\arabic{enumi}}}
\newcommand{\bit}{\begin{itemize}}
\newcommand{\eit}{\end{itemize}}
\newcommand{\bthe}{\begin{theorem}}
\newcommand{\ethe}{\end{theorem}}
\newcommand{\blem}{\begin{lemma}}
\newcommand{\elem}{\end{lemma}}
\newcommand{\bprop}{\begin{proposition}}
\newcommand{\eprop}{\end{proposition}}
\newcommand{\bex}{\begin{example}}
\newcommand{\eex}{\end{example}}
\newcommand{\bas}{\begin{assumption}}
\newcommand{\eas}{\end{assumption}}
\newcommand{\bre}{\begin{remark}}
\newcommand{\ere}{\end{remark}}
\newcommand{\bcor}{\begin{corollary}}
\newcommand{\ecor}{\end{corollary}}
\newcommand{\bdfn}{\begin{definition}}
\newcommand{\edfn}{\end{definition}}

\newcommand{\ones}{\ensuremath{1\!\!1}}

\newcommand{\pset}[1]{\ensuremath{\{#1\}}}
\newcommand{\nset}[1]{\ensuremath{\{1,2,\ldots,#1\}}}

\newcommand{\norm}[1]{\ensuremath{\| #1 \|}}

\newcommand{\R}{\ensuremath{\mathbb R}}

\newcommand{\BP}{\noindent{\bf Proof. }}
\newcommand{\EP}{\hspace*{\fill} $\blacksquare$\bigskip\noindent}
\usepackage[prependcaption,colorinlistoftodos]{todonotes}

\newcommand{\nima}[1]{{\color{black}#1}}
\newcommand{\nimarev}[1]{{\color{black}#1}}
\newcommand{\nimarevtwo}[1]{{\color{black}#1}}
\title{\LARGE \bf Structure-preserving model reduction of physical network systems by clustering}

\author{Nima Monshizadeh and Arjan van der Schaft
\thanks{Nima Monshizadeh and Arjan van der Schaft are with the Johann Bernoulli Institute for Mathematics and Computer Science, University of Groningen, The Netherlands,\newline
{\tt\scriptsize n.monshizadeh@rug.nl}, {\tt\scriptsize a.j.van.der.schaft@rug.nl}}
}

\begin{document}

\maketitle
\thispagestyle{empty}
\pagestyle{empty}

\begin{abstract}
In this paper, we establish a method for model order reduction of a certain class of physical network systems.
The proposed method is based on clustering of the vertices of the underlying graph, and yields a reduced order model within the same class.
To capture the physical properties of the network, we allow for weights associated to both the edges as well as the vertices of the graph.
We extend the notion of almost equitable partitions to this class of graphs.
Consequently, an explicit model reduction error expression in the sense of $\calH_2$-norm is provided for clustering arising from almost equitable partitions.
Finally the method is extended to second-order systems.
\end{abstract}

\section{Introduction}
This paper is a continuation of two preceding papers. In \cite{nimaCONES} an approach was developed for approximation of consensus dynamics by clustering of the vertices of the graph, and error expressions were derived. Furthermore, it was shown in \cite{vdsMTNS14} that by allowing for weights associated to the vertices of the graph one can extend the clustering approach of \cite{nimaCONES} to a structure-preserving model reduction approach, both for first-order and for second-order linear physical network systems.

The main contribution of the present paper is to extend the error expressions for approximation of consensus dynamics obtained in \cite{nimaCONES}  to error expressions for structure-preserving approximation of linear physical network systems. More precisely, while in \cite{nimaCONES} the notion of {\it almost equitable partitions} for unweighted graphs was generalized to graphs with weights on the edges, and an explicit error expression was derived for approximation by clustering according to such an almost equitable partition, we will in this paper further generalize the notion of almost equitable partition to graphs with weights associated both to the edges {\it and} to the vertices, and derive an explicit error expression for model reduction by clustering via such almost equitable partitions. Interestingly, this error expression will depend on the weights associated to the vertices of the graph (but {\it not} on the weights associated to the edges). A direct example is provided by the approximation of {\it mass-damper systems} by clustering of the masses. Finally, we will extend this error expression to model reduction by clustering of second-order systems, as exemplified by {\it mass-spring-damper systems}.

\nimarev{The idea of using clustering for model reduction has been explored before in a number of papers, see e.g. \cite{imuraTAC, Bart, sandberg} and the references quoted therein.
Another perspective for model reduction of multi-agent systems has been taken in \cite{nimaSCL} where the dynamic order of the agents is reduced, yet the interconnection structure remains unchanged.
Nevertheless, the current work differs from other work in at least two aspects: (1) the clustering based model reduction approach proposed in this paper is strictly {\it structure-preserving}, (2) the error expressions are based on the use of {\it almost equitable partitions} of the underlying graph.}
\subsection{Background on graph theory}
An undirected graph $\mathcal{G}(\mathcal{V},\mathcal{E})$, is defined by a set $\mathcal{V}$ of $n$ vertices (nodes) and a set $\mathcal{E}$ of $k$ edges (links, branches), where $\mathcal{E}$ is identified with a set of unordered pairs $\{i,j\}$ of vertices $i,j \in \mathcal{V}$. We allow for {\it multiple} edges between vertices, but not for {\it self-loops} $\{i,i \}$. By endowing the graph with an orientation we obtain a {\it directed graph}. Recall \cite{bollobas} that a directed graph with $n$ vertices and $k$ edges is specified by its {\it incidence matrix}. The incidence matrix, denoted by $D$,  is an $n \times k$ matrix where every column corresponds to an edge of the graph, and contains exactly one $-1$ at the row corresponding to its tail vertex and one $+1$ at the row corresponding to its head vertex, while the other elements are $0$.

Given a directed graph with incidence matrix $D$ and a diagonal positive semi-definite matrix $R$ we can consider the {\it weighted Laplacian matrix} $L:=DRD^T$, where the nonnegative diagonal elements $r_1, \cdots, k,$ of the matrix $R$ are the weights of the edges. It is well-known \cite{bollobas} that this weighted Laplacian matrix $L$ is {\it independent} of the orientation of the graph, and thus is associated with the undirected\footnote{Alternatively \cite{bollobas, godsil} $L$ can be defined as the difference of the diagonal matrix of degrees of the vertices and the adjacency matrix of the undirected graph.} graph $\mathcal{G}$.

Throughout this paper we assume that the graph $\mathcal{G}$ is {\it connected}, or equivalently \cite{bollobas, godsil} $\ker D^T = \spa \ones$ where $\ones$ is the vector of all ones. Otherwise, the proposed model reduction scheme can be applied to the connected components of $\calG$ independently.

\section{Model reduction by clustering for first-order physical network systems}
In this paper we will consider two classes of linear physical network systems; one corresponding to first-order systems on graphs, and the other corresponding to second-order systems. In this section we will concentrate on linear first-order systems; examples of which are mass-damper systems, hydraulic networks, and chemical reaction networks where the substrate and product complexes consist of single chemical species; see e.g. \cite{vdsMTNS14} for further information. The general form of this class of first-order systems is given by
\begin{equation}\label{dyn}
\dot{x} = - DRD^TM^{-1}x + Eu, \quad x \in \mathbb{R}^n, u \in \mathbb{R}^m,
\end{equation}
where $D$ is the incidence matrix of an underlying directed graph $\mathcal{G}$, $R$ is a positive semi-definite matrix, and $M$ is a positive diagonal matrix. Furthermore, $E$ is an $n \times m$ matrix, with every column containing exactly one $1$ element at the row corresponding to the vertex (terminal) where an input applies.
Since the weighted Laplacian matrix $DRD^T$ is independent of the orientation of the graph the system (\ref{dyn}) is, in fact, defined on an undirected graph.

In the case of linear {\it mass-damper systems} (the leading example of this section) the vector $x$ corresponds to the vector of momenta of the $n$ masses associated with the $n$ vertices of the graph, the matrix $M= \diag (m_1, \cdots, m_n)$ is the diagonal $n \times n$ matrix of mass parameters, and $R = \diag (r_1, \cdots, r_k)$ is the diagonal $k \times k$ matrix of damper constants (with every edge corresponding to a damper).
%

Next, we will introduce some abstractions which will come in handy later on, and are important for the interpretation of our approach. The state space of (\ref{dyn}) given by $\mR^n$ can be more abstractly defined as follows; cf. \cite{vdsmaschkeSIAM} for further information. It is given by the linear space $\Lambda_0$ of all functions from the vertex set $\mathcal{V}$ to $\mR$. Obviously $\Lambda_0$ can be identified with $\mR^n$. The matrix $M^{-1}$ defines an inner product on $\Lambda_0$. As a consequence, any vector $M^{-1}x, x \in \Lambda_0$ can be considered to be an element of the dual space of $\Lambda_0$, which is denoted by $\Lambda^0$. For a mass-damper system, $v:= M^{-1}x$ is the vector of velocities of the $n$ masses. It follows that the system (\ref{dyn}) can be also represented in the state vector $v:= M^{-1}x \in \Lambda^0$ as
\begin{equation}\label{dynv}
\dot{v} = - M^{-1}DRD^Tv + M^{-1}Eu, \; v \in \Lambda^0=\mathbb{R}^n, u \in \mathbb{R}^m,
\end{equation}
or equivalently in the gradient system representation
\begin{equation}\label{dynvgradient}
\begin{array}{r}
M\dot{v} = - DRD^Tv + Eu \\
v \in  \Lambda^0= \mathbb{R}^n, u \in \mathbb{R}^m,
\end{array}
\end{equation}
Furthermore, the edge space $\mR^k$ can be defined more abstractly as the linear space $\Lambda_1$ of functions from the edge set $\mathcal{E}$ to $\mR$, with dual space denoted by $\Lambda^1$. It follows that the incidence matrix $D$ defines a linear map (denoted by the same symbol) $D: \Lambda_1 \to \Lambda_0$ with adjoint map $D^T: \Lambda^0 \to \Lambda^1$. Finally, $R$ can be considered to define an inner product on $\Lambda^1$, or equivalently, as a map $R: \Lambda^1 \to \Lambda_1$.
\begin{remark}
Using these abstractions it is straightforward to extend the dynamics (\ref{dyn}) to other spatial domains than just the one-dimensional domain $\mR$. Indeed, for any linear space $\mathcal{R}$ (e.g., $\mathcal{R} = \mR^3$) we can define $\Lambda_0$ as the set of functions from $\mathcal{V}$ to $\mathcal{R}$, and $\Lambda_1$ as the set of functions from $\mathcal{E}$ to $\mathcal{R}$. In this case we can identify $\Lambda_0$ with the tensor product $\mR^n \otimes  \mathcal{R}$ and $\Lambda_1$ with the tensor product $\mR^k \otimes  \mathcal{R}$. Furthermore, the incidence matrix $D$ defines a linear map $D \otimes I : \Lambda_1 \to \Lambda_0$, where $I$ is the identity map on $\mathcal{R}$. In matrix notation $D \otimes I$ equals the Kronecker product of the incidence matrix $D$ and the identity matrix $I$. It is straightforward to extend the formulation of the dynamics (\ref{dyn}) to the case of a general linear space $\mathcal{R}$ instead of $\mR$.
\end{remark}

\nima{As hinted earlier, we interpret the models \eqref{dyn}-\eqref{dynvgradient} as a network of dynamical agents defined on an undirected graph with weighted edges as well as weighted vertices. To incorporate the weights on the edges and vertices at the same time, we define the notion of {\em effective weights} on the vertices as follows.

\begin{definition}
Let $w$ be the weight associated to the edge $\{i,j\}\in \calE$. Also let $\alpha_i \in \R$ and $\alpha_j \in \R$ denote the weights associated to the vertices $i$ and $j$, respectively.
Then, we say that the vertex $i$ receives an {\em effective weight} of $\frac{w}{\alpha_i}$ from the edge $\pset{i,j}$, and, similarly the vertex $j$ receives an effective weight of $\frac{w}{\alpha_j}$.
\end{definition}

Note that in the case of mass-damper system \eqref{dyn}-\eqref{dynvgradient}, the weight of the $i$-th vertex is indicated by the mass parameter $m_i, i=1, \cdots,n.$
The notion of effective weights amounts for normalizing the weights of the edges according to the weights of the corresponding vertices. In other words, effective weights can be interpreted as the (edge) weights received per mass unit by the vertices of the graph.

Following the notion of effective weights, we define the {\em effective Laplacian matrix} as
\beq\label{e:effectiveLaplacian}
L_{\rm eff}:=M^{-1}L=M^{-1}DRD^T
\eeq
It is easy to observe that while the magnitude of the $(i,j)^{th}$ element of the Laplacian matrix $L$ indicates the weight of the edge $\{i,j\}$, the $(i,j)^{th}$ element of $L_{\rm eff}$ represents the effective weight of $\pset{i,j}$ acting on the vertex $i$. Also note that the matrices $L_{\rm eff}^T$ and $L_{\rm eff}$ constitute the state matrices of the systems \eqref{dyn} and \eqref{dynv}, respectively.

\bre
 Note that the matrix $L_{\rm eff}$ is not necessarily symmetric, but it is similar to a symmetric matrix, namely $M^{-\frac{1}{2}}LM^{-\frac{1}{2}}$.
 Of course, one could interpret $L_{\rm eff}$ as the Laplacian matrix of a {\em directed} graph with (nonsymmetric) weights on the edges together with unweighted vertices.
 However, here we prefer the interpretation established above, i.e. an undirected graph with weighted edges and weighted vertices, as it captures more structural/physical properties of the system.
\ere}


Our approach to structure-preserving model reduction of linear physical network systems is based on {\it partitions} of the vertex set of the graph.
Consider a partition of the vertex set $\mathcal{V}$ of the graph $\mathcal{G}$ into $\hat{n}$ disjoint cells $C_1, C_2, \cdots, C_{\hat{n}}$, together with a corresponding $n \times \hat{n}$ {\it characteristic matrix} $P$. The columns of $P$ equal the {\it characteristic vectors} of the cells; the characteristic vector of a cell $C_i$ being defined as the vector with $1$-s at the place of every vertex contained in the cell $C_i$, and $0$ elsewhere.
With a slight abuse of the notation, in what follows, we will denote a partition simply by its characteristic matrix $P$.

As introduced in \cite{vdsMTNS14}, expanding on \cite{nimaCONES}, the basic idea is to {\it reduce} for a given partition $P$ of the graph $\mathcal{G}$, the system (\ref{dyn}) to
\begin{equation}\label{dynred}
\dot{\hat{x}} = - (P^TDRD^TP)(P^TMP)^{-1}\hat{x} + P^TEu
\end{equation}
where $\hat{x}:=P^Tx \in \mathbb{R}^{\hat{n}}$ is the clustered state vector.

We observe that this is again a system of the form (\ref{dyn}). In fact, the matrix $P^TD$ consists of column vectors containing exactly one $+1$ and one $-1$ together with some zero vectors (corresponding to edges which link vertices within a same cell). By leaving out the zero column vectors from $P^TD$ we thus obtain an $\hat{n} \times \hat{k}$ matrix $\hat{D}$, which is the incidence matrix of the {\it reduced graph} $\mathcal{\hat{G}}$, with vertices being the cells of the original graph, and with edges the union of all the edges between vertices in different cells (leaving out edges {\it within} cells). Correspondingly we define $\hat{R}$ as the $\hat{k} \times \hat{k}$ diagonal matrix obtained from $R$ by leaving out the rows and columns corresponding to edges between vertices in a same cell. Finally we define the $\hat{n} \times \hat{n}$ diagonal matrix $\hat{M} = P^TMP$, and $\hat{E} = P^TE$. It follows that the reduced system (\ref{dynred}) is represented as
\begin{equation}\label{dynredfin}
\dot{\hat{x}} = - \hat{D} \hat{R} \hat{D}^T \hat{M}^{-1}\hat{x} + \hat{E}u, \quad \hat{x} \in \mathbb{R}^{\hat{n}}
\end{equation}
For mass-damper systems the dynamics (\ref{dynredfin}) corresponds to the motion of the full-order system where the masses in each cell are rigidly linked to another and move as a single entity with mass equal to the sum of the masses in that cell.

\nima{Note that again the matrix  $(\hat{D} \hat{R} \hat{D}^T\hat{M}^{-1})^T$ can be seen as the effective Laplacian matrix of the reduced order system \eqref{dynredfin}, that is defined on the reduced graph $\hat{\calG}$ with respect to the weights of the edges, given by $\hatR$, as well as weights of the vertices, given by $\hatM$.

Moreover, it is easy to verify that
$$\ker DRD^T=\ker \hatD \hatR \hatD^T,$$
and thus the reduced graph $\hat{\calG}$ is again connected.}

As already indicated in \cite{vdsMTNS14} a Petrov-Galerkin interpretation interpretation of the reduced model (\ref{dynred}) can be given as follows. Recall that a Petrov-Galerkin reduction of a general linear set of differential equations $\dot{x}=Ax$ is given as $\dot{\hat{x}} = W^TAV\hat{x}$, where $V$ and $W$ are matrices of dimension $n \times \hat{n}, \hat{n} < n,$ such that $W^TV=I_{\hat{n}}$. Now the reduced system (\ref{dynred}) is a Petrov-Galerkin reduction of (\ref{dyn}) with $W:=P$, and
\begin{equation}\label{W}
V:= MP\hat{M}^{-1} = MP(P^TMP)^{-1}
\end{equation}
It immediately follows that indeed $W^TV=I_{\hat{n}}$. Furthermore, we note that $W^TW=P^TP$ is a diagonal matrix, and hence $W$ is orthogonal, while also
\[
V^TM^{-1}V= \hat{M}^{-1},
\]
implying orthogonality of $V$ with respect to the inner product defined by $M^{-1}$.


Using the abstractions introduced earlier on we note that the linear maps $V$ and $W$ are actually defined as maps
\[
V: \hat{\Lambda}_0 \to \Lambda_0, \quad W: \hat{\Lambda}^0 \to \Lambda^0
\]
where $\hat{\Lambda}_0$ is the vertex space of the reduced graph, with dual space $\hat{\Lambda}^0$. Moreover, note that $M: \Lambda^0 \to \Lambda_0$ and $\hat{M}: \hat{\Lambda}^0 \to \hat{\Lambda}_0$.

\subsection{Almost equitable partitions and error expressions}
A special case is provided by considering partitions $P$ of the graph $\mathcal{G} (\mathcal{V}, \mathcal{E})$ which are {\it almost equitable} with respect to the weights on the edges defined by the matrix $R$ and the weights on the vertices defined by the matrix $M$. Recall from \cite{cardoso} that a partition of the vertex set $\mathcal{V}$ of an undirected and unweighted graph $\mathcal{G}$ by cells $C_1, C_2, \cdots, C_{\hat{n}}$ with $n \times \hat{n}$ characteristic matrix $P$ is called {\it almost equitable} if for any $p,q \in \{1, \cdots, \hat{n} \}$ with $p \neq q$ there exists an integer $d_{pq}$ such that the number of edges between any vertex of $C_p$ and the cell $C_q$ is equal to $d_{pq}$.
\nima{In other words, the number of neighbors a vertex $i \in C_p$ has in $C_q$ is independent of the choice of $i$, for all $p,q \in \nset{\hat{n}}$ with $p\neq q$.}
It can be shown \cite{cardoso} that a partition $P$ is almost equitable if and only if the subspace $\im P$ is an {\it invariant subspace} of the Laplacian matrix $L:=DD^T$.
In \cite{nimaCONES} the notion of almost equitable partitions was extended to undirected graphs with weights on the edges, and it was shown that a partition $P$ is almost equitable in this sense if and only $\im P$ is invariant with respect to the {\it weighted} Laplacian matrix $DRD^T$. We will further extend this notion to weights both on the edges {\it and} on the vertices.

\begin{definition}\label{d:AEP}
Consider an undirected graph $\mathcal{G}$ with $k$ weighted edges and $n$ weighted vertices. A partition $P$ with cells $C_1, C_2, \cdots, C_{\hat{n}}$ is called {\it almost equitable} if for any $p,q \in \nset{\hat{n}}$ with $p\neq q$ there exists an integer ${w}_{pq}$ such that
$$
\sum_{j \in \calN(i, C_q)} w^i_{\pset{i,j}}=w_{pq}
$$
for all $i \in C_q$. Here, $\calN(i, C_q):=\pset{j \in C_q \mid \pset{i,j} \in \calE}$, and $w^i_{\pset{i,j}}$ denotes the effective weight vertex $i$ receives from the edge $\pset{i,j}$.
\end{definition}

Note that almost equitability in the sense of Definition \ref{d:AEP} means that the sum of the effective weights a vertex $i \in C_p$ receives form the edges located between $C_p$ and $C_q$ is independent of the choice of $i$, for all $p, q$ with $p\neq q$. In the special case where all the vertices have a same weight, this definition boils down to the notion of almost equitability adopted in \cite{nimaCONES}.

We have the following characterization of almost equitability.
\begin{proposition}\label{p:Linv}
Consider an undirected graph $\mathcal{G}$ with weighted edges and weighted vertices. Then a partition of $\mathcal{G}$ with characteristic matrix $P$ is almost equitable if and only if
\begin{equation}\label{e:Linv}
L_{\rm eff} \im P \subset \im P
\end{equation}
where $L_{\rm eff}$ is given by \eqref{e:effectiveLaplacian}.
\end{proposition}
\BP
The proof is analogous to that of \cite[Prop. 1]{cardoso} by adopting the notion of almost equitability in Def. \ref{d:AEP} instead of that introduced in [Sec. 1]\cite{cardoso}.
\EP

%
%
It follows from standard linear algebra that given an almost equitable partition with respect to $R$ and $M$ given by the $n \times \hat{n}$ characteristic matrix $P$ we can construct an
$n \times (n - \hat{n})$ orthogonal matrix $S$ such that $\im S$ is orthogonal to $M \im P$ and the $n \times n$ stacked matrix $\begin{bmatrix} P  & S \end{bmatrix}$ is invertible. By \eqref{e:Linv}, we obtain that
\begin{multline*}
 \begin{bmatrix} P^T \\ S^T \end{bmatrix} DRD^T  \begin{bmatrix} P & S \end{bmatrix} \\
 =\begin{bmatrix} P^TDRD^TP & 0 \\ 0 & S^TDRD^TS
 \end{bmatrix}
\end{multline*}

Note that the inverse of the transpose of $\begin{bmatrix} P  & S \end{bmatrix}$ is determined by
 \[
 \begin{bmatrix} P^T \\ S^T \end{bmatrix}
 \begin{bmatrix} MP(P^TMP)^{-1} & MS(S^TMS)^{-1} \end{bmatrix} = I_n
 \]
It follows that the state transformation
\[
z := \begin{bmatrix} P^T \\ S^T \end{bmatrix} x
\]
transforms the matrix $DRD^TM^{-1}$ into
{\small
\begin{align*}
&\begin{bmatrix} P^T \\ S^T \end{bmatrix} DRD^TM^{-1} \begin{bmatrix} MP(P^TMP)^{-1} &  MS(S^TMS)^{-1} \end{bmatrix}  \\
&\quad=
\begin{bmatrix} \hat{D} \hat{R} \hat{D}^T & 0 \\
0 &  S^TDRD^TS \end{bmatrix} \begin{bmatrix} \hat{M}^{-1} & 0 \\ 0 & (S^TMS)^{-1} \end{bmatrix}
\end{align*}}
where we have used \eqref{e:Linv} to obtain the first equality, and the matrices $\hat{D}, \hat{R}, \hat{M}$ are defined as before. Hence the dynamics for the transformed state
\begin{equation}
\nonumber
z :=\begin{bmatrix} P^T \\ S^T \end{bmatrix} x =: \begin{bmatrix} \hat{x} \\ x' \end{bmatrix}
\end{equation}
completely decouples into a dynamics for $\hat{x}$ given by
\begin{equation}\label{e:red-AEP}
\dot{\hat{x}} = - \hat{D} \hat{R} \hat{D}^T \hat{M}^{-1}\hat{x} + P^TEu,
\end{equation}
which is the reduced model as proposed before, together with the remaining dynamics
\begin{equation*}
\dot{x}'  = - S^TDRD^TS (S^TMS)^{-1} x' +S^TEu,
\end{equation*}
which can be regarded as an error dynamics.

\nima{
Observe that that internal dynamics of \eqref{dyn}-\eqref{dynvgradient} is governed by the differences of the velocity of the masses given by $D^T M^{-1}x$ (equivalently, $D^Tv$).
Of course, it is important that we have a {\em valid} approximation of these variables in the proposed reduced model.
To this end, we define output variables $y$ as
\beq\label{e:y}
y=R^{\frac{1}{2}}D^\top M^{-1}x=R^{\frac{1}{2}}D^Tv
\eeq
Note that we have
$$\norm{y}_2^2=v^TDRD^T v= - \frac{d \frac{1}{2}v^\top M v}{dt}$$
which can be interpreted as the power dissipated by the dampers.

Then the output variables of the reduced order model are given by
\beq\label{e:y-reduced}
y=\hat{R}^{\frac{1}{2}}\hatD^\top \hatM^{-1}\hat{x}=\hat{R}^{\frac{1}{2}}\hatD^T\hat{v}
\eeq
where the matrices $\hatR$, $\hatD$, $\hatM$ are defined as before, and $\hat{v}:=M^{-1}\hat{x}$. Now, let $G$ denote the transfer matrix from $u$ to $y$ in the original model \eqref{dyn}
(equivalently, \eqref{dynv}) with the output variables \eqref{e:y}. Then, it is well-known that
$$\norm{G}_2^2=\trace E^TM^{-1}XM^{-1}E$$
where $\norm{G}_2$ denotes the $\calH_2$-norm of $G$ and $X$ is given by
$$
X=
\int_{0}^{\infty} e^{-LM^{-1}t}L e^{-M^{-1}Lt} dt
$$
with $L=DRD^T$.
It can be verified that
\begin{multline*}
e^{-LM^{-1}t}L e^{-M^{-1}Lt}dt\\
=d(-\frac{1}{2}e^{-LM^{-1}t}M e^{-M^{-1}Lt}).
\end{multline*}
Hence, we have
\beq\label{e:Xsol}
X=-\frac{1}{2}e^{-LM^{-1}t}M e^{-M^{-1}Lt}\mid _0^\infty
\eeq

Moreover, as the matrix $M^{-1}L$ is diagonalizable, we have
$$e^{-M^{-1}Lt}=\sum_{i=1}^n \mathbf{v}_i \mathbf{w}_i^T e^{-\lambda_it}$$
where $\lambda_i$, $\mathbf{v}_i$, and $\mathbf{w}_i, i=1, \cdots,n,$  are the eigenvalues, right eigenvectors, and left eigenvectors of $M^{-1}L$ respectively.
Moreover, here the eigenvectors are chosen such that $\mathbf{w}_i^T\mathbf{v}_j$ is equal to $1$ for $i=j$, and is equal to $0$ otherwise.
Due to the connectivity of $\calG$, the matrix $M^{-1}L$ has a single eigenvalue at zero, say $\lambda_1=0$, and the rest of the eigenvalues are real and strictly positive.
In addition, the corresponding left and right eigenvectors of $\lambda_1$ can be chosen as $\mathbf{w}_1=\frac{1}{\sigma_M}M\ones$ and $\mathbf{v}_1=\ones$, where $\sigma_M$ denotes the sum of all the masses, i.e. $\sum_i m_i$, and  $\ones$ denotes the vector of ones with an appropriate dimension.  Hence, we have
\footnote{Note that \eqref{e:lim} guarantees that $X$ is well-defined, as $\underset{t \rightarrow \infty  }{\lim} Le^{-M^{-1}Lt}=0$.}
\beq\label{e:lim}
\lim_{t \rightarrow \infty  } e^{-M^{-1}Lt}=\frac{1}{\sigma_M}\ones\ones^\top M.
\eeq
Therefore, by \eqref{e:Xsol}, the matrix $X$ is computed as
\beq\label{e:Xfinal}
X=\frac{1}{2}(M-\frac{1}{\sigma_M}M\ones \ones^TM)
\eeq

Consequently, we obtain that
\begin{align}\label{e:normG}
\nonumber \norm{G}_2^2&=\frac{1}{2}\trace E^T(M^{-1}-\frac{1}{\sigma_M}\ones \ones^T)E\\
\nonumber &=\frac{1}{2}\trace EE^T(M^{-1}-\frac{1}{\sigma_M}\ones \ones^T)\\
&=\frac{1}{2}\sum_{i\in V_f}(\frac{1}{m_i}-\frac{1}{\sigma_M})
\end{align}
where $V_f$ denotes the set of {\em forced vertices}, i.e. vertices to which inputs (forces) are applied. Note that the set
$V_f$ is identified by the nonzero elements of the diagonal matrix $EE^T$.

Now, let $\norm{\hatG}_2$ denote the $\calH_2$-norm of the transfer matrix from $u$ to $y$ in the reduced order model \eqref{e:red-AEP} together with the output variables \eqref{e:y-reduced}.
In a similar fashion to the derivation of $\norm{G}_2^2$, $\norm{\hatG}_2^2$ can be computed as
\beq\label{e:normGhat}
\norm{\hatG}_2^2=\frac{1}{2}\trace \hatE \hatE ^T(\hatM^{-1}-\frac{1}{\sigma_{\hatM}}\ones \ones^T)
\eeq
with $\sigma_{\hatM}=\sigma_M$ as the total sum of the masses is invariant under reduction.

\nimarevtwo{Let $r_i$ be an integer such that $i\in C_{r_i}$ for each $i\in \calV$.
By $\sigma_M^i$, we denote the sum of the masses which belong to the same cell as $i$, i.e.
$$\sigma_M^i:=\sum_{j\in C_{r_i}}m_j,$$
for each $i\in \calV$.}
Then, \eqref{e:normGhat} amounts to
\beq\label{e:normGhat1}
\norm{\hatG}_2^2=\frac{1}{2}\sum_{i\in V_f}(\frac{1}{\sigma_M^i}-\frac{1}{\sigma_M}).
\eeq

From the derivation of \eqref{e:red-AEP} and by Proposition \ref{p:Linv}, it is easy to verify that
$$\norm{G}_2^2=\norm{G-\hatG}_2^2+\norm{\hatG}_2^2.$$
Therefore, by \eqref{e:normG} and \eqref{e:normGhat1}, we obtain that
$$
\norm{G-\hatG}_2^2=\frac{1}{2}\sum_{i\in V_f}(\frac{1}{m_i}-\frac{1}{\sigma_M^i}).
$$
which corresponds to the model reduction error associated to the partition $P$.
This is summarized in the following theorem.
}
\nima{
\begin{theorem}
Let $G$ denote the transfer matrix from $u$ to $y$ in \eqref{dyn} with output variables \eqref{e:y}.
Assume that $P$ is an almost equitable partition of $L_{\rm eff}$, with $L_{\rm eff}$ given by \eqref{e:effectiveLaplacian}, in the sense of Definition \ref{d:AEP}.
Corresponding to $P$, let $\hatG$ denote the transfer matrix from $u$ to $y$  in the reduced
order model \eqref{dynred} with output variables \eqref{e:y-reduced}.
By $\Xi:=\norm{G-\hatG}_2^2$, we denote the model reduction error associated with the partition $P$. Then, we have
\beq\label{e:error}
\Xi=\frac{1}{2}\sum_{i\in V_f}(\frac{1}{m_i}-\frac{1}{\sigma_M^i})
\eeq
where $V_f$, $m_i$ and $\sigma_M^i$ are defined as before.
\end{theorem}
The formula \eqref{e:error} indicates that in order to have a small model reduction error the masses of the forced vertices, i.e. the vertices in $V_f$, should be the dominant masses of their corresponding clusters.  That is,  $m_i$ should be close to  $\sigma_M^i$ for $i\in V_f$. An interesting special case is obtained when $m_i=\sigma_M^i$, meaning that the forced vertices appear as singleton in $P$. This clearly results in $\Xi=0$. In fact, in this case model reduction by the proposed clustering scheme amounts to removing the uncontrollable modes of the system, thus keeping the input-output behavior of the system unchanged.

}
\bigskip{}

\section{Second-order physical network systems}
The leading example of this section will be {\it mass-spring-damper systems} on graphs \cite{vdsmaschkeSIAM}. As in the case of mass-damper systems the masses are associated to the vertices of a graph $\mathcal{G}$, but now part of the edges corresponds to the {\it springs} and the other part of the edges is associated with the {\it dampers}. This means that the incidence matrix of the graph can be split into an incidence matrix $D_s$ corresponding to the spring edges and an incidence matrix $D_d$ corresponding to the dampers. Then \cite{vdsmaschkeSIAM} the dynamics of the mass-spring-damper system takes the port-Hamiltonian form
\begin{equation}\label{dyn2PH}
\begin{bmatrix} \dot{q} \\ \dot{p} \end{bmatrix} =
\begin{bmatrix} 0 & D_s^T \\ -D_s & -D_dRD_d^T \end{bmatrix}
\begin{bmatrix} Kq \\ M^{-1}p \end{bmatrix}+
\nimarev{\bbm 0\\E\ebm u}
\end{equation}
Here $p \in \mathbb{R}^n$ is a vector of momentum variables associated to the vertices of the graph (similar to the vector denoted before by $x$), and $q \in \mathbb{R}^k$ is a vector of state variables associated to the $k$ edges of the graph, modeling the elongations of the springs. \nimarev{The matrices $M$ and $E$ are defined as before. Also the matrices $R$ and $K$ are positive semi-definite diagonal matrices indicating the weights of the edges corresponding to the dampers and the springs, respectively.}  Furthermore, the total energy (Hamiltonian) is given as
\[
H(q, p) = \frac{1}{2}p^TM^{-1}p + \frac{1}{2}q^TKq
\]
Equivalently, this can be written out as a second-order system in the vector of state variables $p$ associated to the vertices as
\begin{align*}
\ddot{p} &= - D_dRD_d^TM^{-1}\dot{p} + D_sK\dot{q}\nimarev{+Eu} \\
&=- D_dRD_d^TM^{-1}\dot{p} - D_sKD_s^TM^{-1}p\nimarev{+Eu}
\end{align*}
or equivalently
\begin{equation}\label{dyn2}
\ddot{p} + D_dRD_d^TM^{-1}\dot{p} + D_sKD_s^TM^{-1}p =\nimarev{Eu}
\end{equation}
Note that since the weighted Laplacian matrices $D_dRD_d^T$ and $D_sKD_s^T$ are independent of the orientation of the graph, the second-order system (\ref{dyn2}) is in fact defined on the {\it undirected} graph $\mathcal{G}$.

\smallskip
Now consider any partition $P$ for the graph $\mathcal{G}$. Then we define a reduced model for (\ref{dyn2}) as the following second-order system with clustered vector $\hat{p} :=P^Tp$ given as
\begin{equation}\label{dyn2red}
\ddot{\hat{p}} + \hat{D}_d\hat{R}\hat{D}_d^T\hat{M}^{-1}\dot{\hat{p}} + \hat{D}_s\hat{K}\hat{D}_s^T\hat{M}^{-1}\hat{p} =\nimarev{\hatE u},
\end{equation}
where $\hat{D}_d, \hat{D}_b$ are defined similarly as $\hat{D}$ before. \nimarev{Likewise, the matrices $\hatE$ and $\hat{R}$ are defined as before. In the same vein, the matrix} $\hat{K}$ is the diagonal matrix obtained from $K$ by leaving out the rows and columns corresponding to the edges {\it within} any of the cells.

A first-order state space formulation of the reduced system (\ref{dyn2red}) is given as the port-Hamiltonian system on the reduced graph
\begin{equation}\label{dyn2PHred}
\begin{bmatrix} \dot{\hat{q}} \\ \dot{\hat{p}} \end{bmatrix} =
\begin{bmatrix} 0 & \hat{D}_s^T \\ \hat{D}_s & -\hat{D}_d\hat{R}\hat{D}_d^T \end{bmatrix}
\begin{bmatrix} \hat{K}\hat{q} \\ \hat{M}^{-1}p \end{bmatrix}\nimarev{+\bbm 0\\\hatE\ebm u}
\end{equation}
where $\hat{q}$ corresponds to the edges which survive in the reduced model.

\nima{
Similar to \eqref{e:effectiveLaplacian}, one can define the effective Laplacian with respect to the edges corresponding to the dampers as
$$L_{\rm eff}^d:=M^{-1}D_dRD_d^T,$$
and with respect to the edges corresponding to the springs as
$$L_{\rm eff}^s:=M^{-1}D_sKD_s^T.$$

Explicit error expressions can be again provided in the case of a partition $P$, which is almost equitable with respect to both $L_{\rm eff}^d$ and $L_{\rm eff}^s$ .
That is we need $\im P$ to satisfy
\begin{equation}
L_{\rm eff}^d \im P \subset \im P, \quad L_{\rm eff}^s \im P \subset \im P
\end{equation}
It can be shown that by taking output variables as in \eqref{e:y} with $D=D_d$, the expression for the reduction error will be the same as \eqref{e:error}. We leave the verification to the readers.
}

\section{Conclusions and Outlook}
We have established a model reduction method for a class of physical network systems.
The network topology has been given by an undirected graph with weighted edges as well as weighted vertices. Reduced order models within the same class of systems have been obtained by clustering the vertices of the underlying graph. We have introduced the notion of effective weights and effective Laplacian matrix in order to extend the ordinary definition of almost equitable partitions to the class of graphs with weighted vertices. In the case that the clustering decision correspond to an almost equitable partition, an explicit model reduction error in the sense of $\calH_2$-norm is provided.  Main questions under current consideration are the approximation properties of the reduced system depending on the choice of the partition, {\nima{including partitions which are not almost equitable}, as well as the relations to model reduction by taking Schur complements of the weighted Laplacian of the graph; see e.g. \cite{vdsraojaya} and the references quoted therein. Another question concerns the extension to the nonlinear case.


\end{document}